\documentclass[conference]{IEEEtran}

\usepackage{graphicx}
\usepackage{amsfonts}
\usepackage{multirow}
\usepackage{makecell}
\usepackage{subfigure}
\usepackage[numbers]{natbib} 

\usepackage{color}

\def\BibTeX{{\rm B\kern-.05em{\sc i\kern-.025em b}\kern-.08em
    T\kern-.1667em\lower.7ex\hbox{E}\kern-.125emX}}

\begin{document}

\title{Social Influence Prediction with Train and Test Time Augmentation for Graph Neural Networks}

\makeatletter
\newcommand{\linebreakand}{%
  \end{@IEEEauthorhalign}
  \hfill\mbox{}\par
  \mbox{}\hfill\begin{@IEEEauthorhalign}
}
\makeatother

\author{\IEEEauthorblockN{ Hongbo Bo}
\IEEEauthorblockA{\textit{Department of Computer Science} \\
\textit{University of Bristol}\\
Bristol, UK\\
hongbo.bo@bristol.ac.uk}
\and
\IEEEauthorblockN{ Ryan McConville}
\IEEEauthorblockA{\textit{Department of Engineering Mathematics} \\
\textit{University of Bristol}\\
Bristol, UK\\
ryan.mcconville@bristol.ac.uk}
\linebreakand
\IEEEauthorblockN{Jun Hong}
\IEEEauthorblockA{\textit{Department of Computer Science} \\ \textit{ and Creative Technologies} \\
\textit{University of the West of England}\\
Bristol, UK\\
Jun.Hong@uwe.ac.uk}
\and
\IEEEauthorblockN{Weiru Liu}
\IEEEauthorblockA{\textit{Department of Engineering Mathematics} \\
\textit{University of Bristol}\\
Bristol, UK\\
weiru.liu@bristol.ac.uk}}

\maketitle

\begin{abstract}
   Data augmentation has been widely used in machine learning for natural language processing and computer vision tasks to improve model performance. However, little research has studied data augmentation on graph neural networks, particularly using augmentation at both train- and test-time. Inspired by the success of augmentation in other domains, we have designed a method for social influence prediction using graph neural networks with train- and test-time augmentation, which can effectively generate multiple augmented graphs for social networks by utilising a variational graph autoencoder in both scenarios. We have evaluated the performance of our method on predicting user influence on multiple social network datasets. Our experimental results show that our end-to-end approach, which jointly trains a graph autoencoder and social influence behaviour classification network, can outperform state-of-the-art approaches, demonstrating the effectiveness of train- and test-time augmentation on graph neural networks for social influence prediction. We observe that this is particularly effective on smaller graphs.
\end{abstract}


\section{Introduction}
Graph neural networks (GNNs)~\cite{scarselli2008graph} have been shown to be effective in various graph machine learning tasks, such as link prediction and node classification. The rapid growth of online social networks has led to the development of numerous methods for studying social behaviour online. However, many learning tasks on social networks have relied heavily on manual feature extraction. GNNs have provided an alternative  to this with their ability to automatically learn representations end-to-end.  One such task of interest, which has been shown to be enhanced using GNNs, is social influence prediction~\cite{qiu2018deepinf}.

Data augmentation~\cite{shorten2019survey}, which increases the amount of data available by creating informative variations of existing data, can improve the performance of machine learning models and has been widely used in many machine learning tasks~\cite{zhao2020data, hoffer2019augment, xie2019unsupervised}. In the fields of computer vision (CV)~\cite{zhao2019data, shorten2019survey} and natural language processing (NLP)~\cite{yoo2019data}, the combination of data augmentation methods and deep neural networks have been shown to be effective. By performing data augmentation, model performance can be improved as it facilitates the neural network to learn generalizable features related to the task. While GNNs have become a popular research field, little research has focused on using data augmentation for GNNs, especially in terms of using augmentation at both train- and test-time. Motivated by the success of data augmentation in CV and NLP, we study whether data augmentation at not only train-time, but also test-time, can improve the performance of GNNs, particularly on the task of social influence prediction.

Extending the work of DeepInf~\cite{qiu2018deepinf}, we have developed a method, AugInf, for social influence prediction with both train- and test-time augmentation for GNNs. In this method, the augmented graphs are first generated by utilising a variational graph autoencoder (VGAE)~\cite{kipf2016variational} and then social influence is predicted by joint training of both a Graph Auto-Encoder (GAE)~\cite{kipf2016variational} and a GNN prediction module based on either a Graph Convolutional Network (GCN)~\cite{kipf2017semi} or a Graph Attention Network (GAT)~\cite{velivckovic2017graph}. We have compared the performance of AugInf with several state-of-the-art GNN approaches by experimenting on numerous social networks. Our experimental results show that AugInf can improve prediction performance on several of these social networks.

In summary, our contributions are as follows. First, we propose a joint training approach consisting of Graph Auto-Encoder (GAE) ~\cite{kipf2016variational} and GNN prediction module.
The GNN prediction module is implemented as either a Graph Attention Network (GAT)~\cite{velivckovic2017graph} or a Graph Convolutional Network (GCN)~\cite{kipf2017semi}.
The joint training approach optimizes for two tasks simultaneously. The first is to obtain more effective latent representations of input graphs, while the second is to utilize the resulting representations to improve social influence predictive performance. Additionally, we use the augmentation approach at both train- and test-time to further improve the joint training model performance, which we demonstrate with an ablation study. To the best of our knowledge, we are the first to explore test-time augmentation on GNNs, and therefore the first to combine both train-time and test-time augmentation on GNNs. Finally, we conduct an experimental evaluation on multiple different social networks comparing our proposed method with two state-of-the-art methods.

\section{Related Work}

\paragraph{Graph Neural Networks} Graph Neural Networks (GNNs)~\cite{zhang2020deep, wu2020comprehensive} have rapidly grown to become a popular research area, providing a highly competitive approach for tasks involving graph data. One line of research focuses on unsupervised models, e.g. VGAE~\cite{kipf2016variational} and Graphite~\cite{grover2019graphite}.  These unsupervised variational models typically aim to use generative modelling of graphs for graph reconstruction, link prediction and clustering. Additionally, supervised models have attracted significant attention, such as SCNN~\cite{bruna2013spectral}, ChebyNet~\cite{defferrard2016convolutional}, GAT~\cite{velivckovic2017graph} and GCN~\cite{kipf2017semi}, which are widely used in tasks where labelled data is available.

\paragraph{Data Augmentation} Data augmentation has been shown to be an effective approach in machine learning which expands a dataset by producing transformed copies of data, thereby making the model invariant to these transformations. Data augmentation has been widely used to improve generalizability of machine learning models in natural language processing (NLP) and computer vision (CV). Most of the work on data augmentation has focused on improving augmentation at the training phase, e.g., batch augmentation~\cite{hoffer2019augment} and UDA~\cite{xie2019unsupervised}. There are also studies that focus on augmentation during the testing phase~\cite{he2015delving}. However, data augmentation for graph neural networks has only been recently studied, such as  SUBG-CON~\cite{jiao2020sub} and NodeAug~\cite{wang2020nodeaug}. Particularly, there is no research on test-time augmentation for GNNs.

\paragraph{Social Influence}
The Independent Cascade Model~\cite{saito2008prediction} and Linear Threshold Model~\cite{chen2010scalable} are classic social network influence propagation models.
Measuring social influence can be broadly divided into two categories based on the methods used. The first category typically utilizes ranking algorithms such as TwitterRank~\cite{weng2010twitterrank},  Truetop~\cite{zhang2015truetop} and EIRank~\cite{bo2020social}, to quantify each user’s influence. This category of methods provide a coarse influence value for each user at a specific time-point and do not model changes to influence, nor the direct effect of this influence on others on the network. The second category of methods are based on predictive models to estimate social influence change. For example, on the global-level patterns of social influence, the DeepCas model~\cite{li2017deepcas} can predict the information cascade by using recurrent neural networks. DeepInf~\cite{qiu2018deepinf} and NNMLInf~\cite{wang2019nnmlinf} consider a social influence prediction task as a label classification task and directly predict the user behaviour with respect to influence.

\section{Preliminaries}
Let $G=(V,E)$ be an input graph which consists of a set of nodes $V$ and a set of edges $E$, where $E\in V\times V$. In the most general sense, a social network can be represented by a graph $G$, where $V$ represents users and $E$ is the set of directed edges representing how the users are connected. The adjacency matrix $A$ is a (0,1) matrix with 0s on its diagonal (i.e. no self-connection) which can represent graph $G$, where $A_{ij} = 0$ indicates nodes $i$ and $j$ are not connected. If $G$ is undirected, $A_{ij} = 1$ indicates nodes $i$ and $j$ are connected and $A$ is symmetric. If $G$ is directed, $A_{ij} = 1$ indicates there is a link from node $i$ to node $j$. A graph neural network model is defined as $f(X, A)$, where $X$ is the node feature matrix of $A$.

\subsection{Graph Convolutional Network}
A Graph Convolutional Network (GCN)~\cite{kipf2017semi} is a semi-supervised learning algorithm for graph data, typically used for node and graph classification, as well as link prediction. A GCN model is typically formed by stacking multiple GCN layers, and for each GCN layer, the inputs are the adjacency matrix $A$ and the features matrix, $H\in\mathbb{R}^{n\times F}$, where $n$ is the number of vertices, and $F$ is the number of features. 
For each GCN layer:
\begin{equation}
    H^{(l+1)}=\sigma(\widetilde D^{-1/2}\widetilde A\widetilde D^{1/2}H^{(l)}W^{(l)}),
\label{gcn}\end{equation}
where $\widetilde{A}$ is the adjacency matrix $A$ with added self-connections (the diagonal elements of the matrix are 1), $\widetilde{D}$ is the diagonal degree matrix where ${\widetilde D}_{ii}={\textstyle\sum_j}{\widetilde A}_{ij}$, and $\sigma(\cdot)$ denotes an activation function. The input layer $H^{(0)}=X$.
We will experiment with the use of a GCN model for social influence prediction.

\subsection{Graph Attention Network} 
A Graph Attention Network (GAT)~\cite{velivckovic2017graph} is an attention-based version of GCN, which incorporates self-attention mechanisms. The GAT layer performs the self-attention mechanism for each node by introducing attention coefficients. The self-attention mechanism is an attention function $attn$:
\begin{equation}
    \mathbb{R}^{F^\prime}\times\mathbb{R}^{F^\prime}\rightarrow{\mathbb{R}},
\end{equation}
where $F^\prime$ represents the number of output features of each node in a GAT layer and the attention coefficient between each pair of nodes is calculated as:
\begin{equation}
    e_{ij}=attn(Wh_i,Wh_j),
\end{equation}
where $h_i$, $h_j \in \mathbb{R}^{F}$ and $W \in \mathbb{R}^{F^\prime} \times \mathbb{R}^{F}$, and the attention function $attn$ is instantiated with a dot product and a LeakyReLU~\cite{xu2015empirical} non-linearity. The attention coefficient $e_{ij}$ is considered as the importance of node $j$ to node $i$. A \textit{softmax} function is adopted to the normalized attention coefficient to make it easier to calculate coefficients and compare them among nodes:
\begin{equation}
    \alpha_{ij}=softmax_j(e_{ij})={\textstyle\frac{Exp(e_{ij})}{\sum_{k\in N_i}Exp(e_{ik})}},
\end{equation}
where $N_i$ is the set of neighbour nodes of node $i$ and $\alpha_{ij}$ is the coefficient for aggregating the calculation of the output feature ${h_i}^\prime$ which has incorporated neighborhood information:
\begin{equation}
    {h_i}^\prime=\sigma({\sum_{j\in N_i}}\alpha_{ij}Wh_j),
\label{hi}\end{equation}
where $\sigma$ is a non-linear function. In order to make the self-attention learning process more stable, it is  known to be effective to use multi-head attention to expand the attention mechanism. By using $K$ independent attention mechanisms to perform transformations with Equation~\ref{hi}, and then concatenating their features together, the following output can be obtained:
\begin{equation}
    h_i^\prime=\parallel_{k=1}^K\sigma(\sum_{j\in N_i}\alpha_{ij}^kWh_j^k),
\end{equation}
where $\parallel$ denotes the vector concatenation operation and the dimension of $h_i^\prime$ is $KF^\prime$.
We will experiment with the use of a GAT model for social influence prediction.

\subsection{Graph Auto-Encoder}
A Graph Auto-Encoder (GAE)~\cite{kipf2016variational} can utilize the GCN layers to obtain the latent representations of the nodes in the graph through an encoder-decoder structure to learn representations for downstream tasks, such as link prediction and node classification. The encoder process can be expressed as:
\begin{equation}
    Z=GCN(X,A), 
    \label{Z}
\end{equation}
where $Z\in\mathbb{R}^{n\times F}$ is the latent representation, which is also referred to as an embedding. In this paper, we use a two layer GCN autoencoder which can be defined as:
\begin{equation}
    GCN(X,A)=\widetilde A\sigma(\widetilde AXW_0)W_1,
    \label{GCN} 
\end{equation}
where $W_0$ and $W_1$ are the parameters to be learned and $\sigma$ is the ReLU activation function. GAE uses the inner-product as a decoder to reconstruct the original graph. In the training process of GAEs, cross entropy is used as the loss function:
\begin{equation}
    \mathcal L=\frac1N\sum_{i=1}^N-(y_i\log{\widehat{y}_i}+(1-y_i)\log(1-{\widehat{y}_i}))
\label{loss}\end{equation}
In the above equation, $y_i$ represents the value of an element in the adjacency matrix $A$ (0 or 1), and $N$ is the size of $A$, $\widehat{y}_i$ represents the value of the corresponding element in the reconstructed adjacency matrix $\widehat{A}$ (between 0 and 1). We use GAE in the joint training model to obtain the latent representation.

\subsection{Variational Graph Auto-Encoder}
A Variational Graph Auto-Encoder (VGAE)~\cite{kipf2016variational} uses latent variables for the model to learn the distributions, and then samples from these distributions to get latent representations. In VGAE, $Z$ is no longer obtained by a certain function (such as eq. (\ref{Z})), but by sampling from a Gaussian distribution. This mechanism makes VGAE suitable for graph generation tasks. VGAE uses a 2-layer GCN model (same as Equation. (\ref{GCN})) to calculate the mean $\mu$ and variance $\sigma$ respectively to obtain a Gaussian distribution:
\begin{equation}
    \mu=GCN_\mu(X,A),
\end{equation}
\begin{equation}
    \log_\sigma=GCN_\sigma(X,A),
\end{equation}
$W_0$ is the same in both $GCN_\mu$ and $GCN_\sigma$, but $W_1$ is different. $Z$ can be calculated by \textit{reparameterization}~\cite{kingma2013auto} and the decoder of VGAE is also an inner-product. The loss function of VGAE is:
\begin{equation}
    \mathcal L={\mathcal L}_{CE}+{\mathcal L}_{KLD},
\end{equation}
Where ${\mathcal L}_{CE}$ is same as Equation~\ref{loss} and ${\mathcal L}_{KLD}$ is the KL divergence which is $\frac12\sum\nolimits_{i=1}^d(\sigma_i^2+\mu_i^2-\log(\sigma_i^2)-1)$, where $d$ is the dimension of $Z$. We use VGAE as part of the graph augmentation process.

\section{Method}
We propose a method for social influence prediction which consists of a joint training of the GAE and the GNN prediction module with both train- and test-time augmentation. For clarity, as different strategies are used for train- and test-time augmentation, we demonstrate the entire method in Figure~\ref{method} with shared boxes showing training and testing stages, respectively.

\begin{figure*}[]
\centering
  \includegraphics[width=\textwidth]{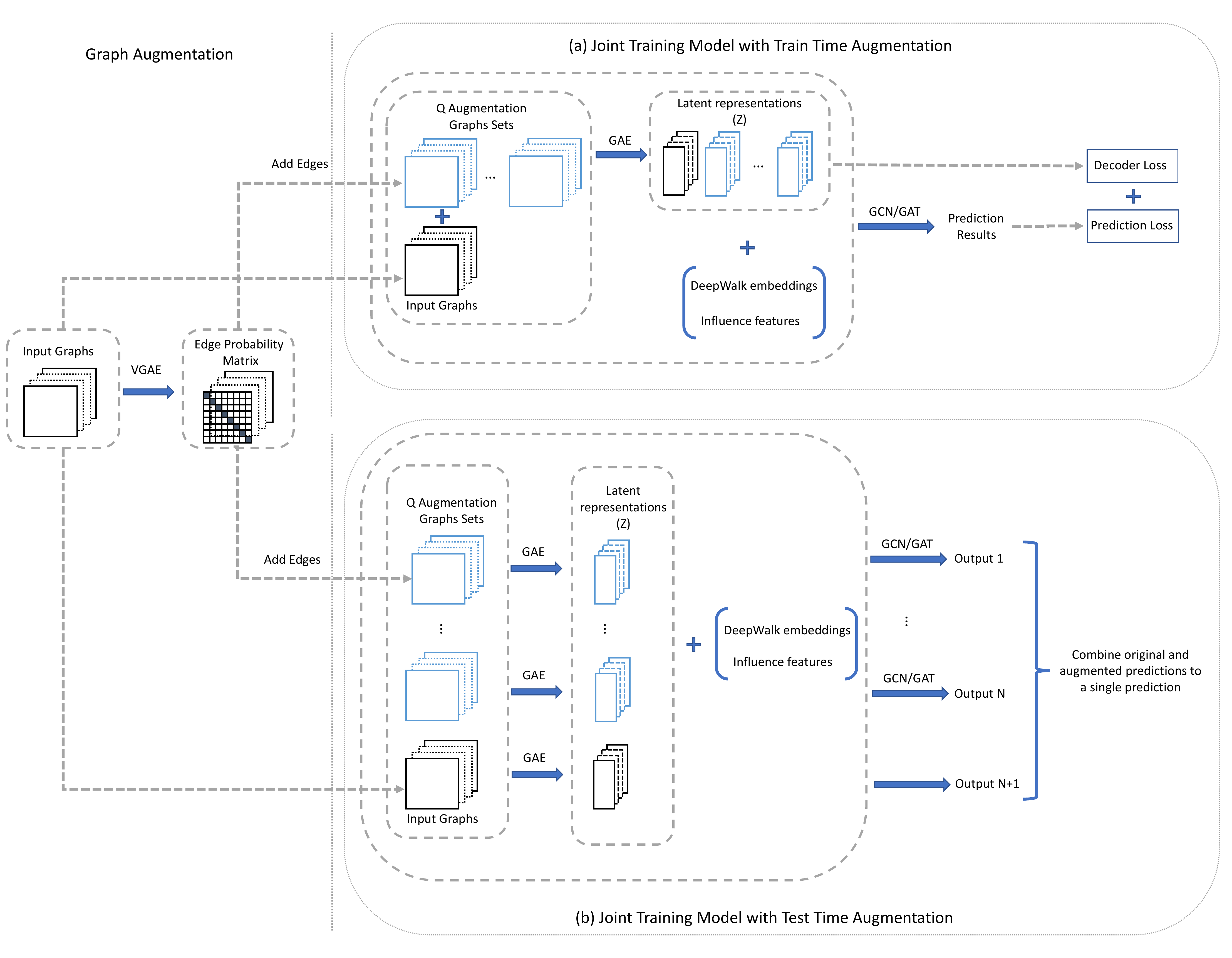} 
  \caption{The AugInf method. AugInf will first obtain the predicted edge probability matrix $M$ by using the VGAE~\cite{kipf2016variational}. A threshold hyperparameter will control the amount of edges added during the augmentation. We extend the work of~\cite{zhao2020data} to perform multiple augmentations, randomly sampling a subset of edges to add to each original subgraph. (a) For train-time augmentation, the method will then integrate all augmentations together for the joint training process. The joint training model has two steps:  the GAE step will learn the latent representations of the input data for the prediction stage; the prediction module (GAT or GCN) will produce the social influence predictions. The losses of these two stages (decoder loss and prediction loss) will be combined and propagated backwards to jointly update the model. (b) For test-time augmentation, AugInf will generate several augmentations of each test (validation) example, learning a representation for each before producing the influence prediction. AugInf will calculate the average of the predictions to produce the final social influence prediction.}
  \label{method} 
\end{figure*}

\subsection{Data Augmentation}
 Before the joint training, we first obtain augmentations of our data by adapting a similar idea to ~\cite{zhao2020data}. The approach for augmentation consists of two steps: (1) using the variational graph auto-encoder (VGAE)~\cite{kipf2016variational} to obtain edge probabilities for all possible and existing edges in graph  $G$, (2) using the predicted edge probabilities, with a threshold set to stochastically add new edges, creating a modified graph $G_m$, which is used as input to the joint training process. The key idea of this augmentation approach is to use information inherent in the graph to predict which non-existent edges are likely to be added to the augmented graph to improve generalization. As a VGAE is a generative model, we utilize VGAE as part of our graph augmentation process.

The VGAE consists of a two-layer GCN encoder and an inner-product decoder:
\begin{equation}
    M=\sigma(ZZ^T),
\end{equation}
where $Z$ is the latent representation, $\sigma$ is an element-wise sigmoid function and $M$ is the predicted (symmetric) edge probability matrix produced by the inner-product decoder. We can add edges in graph $G$ to obtain the augmentation graph $G_m$ according to the probabilities in $M$. Corresponding to $G=(V,E)$, $G_m$  can be expressed as $G_m=(V,E_m)$, and for each node $v\in V$, there are different edge sets $e$ and $e_m$ in $G$ and $G_m$ respectively.

For efficiency, the input graph we will use is not an entire graph $G$, but a set of subgraphs sampled by random walks on $G$, resulting in  $N$ subgraphs, following the approach of DeepInf~\cite{qiu2018deepinf}. We will perform augmentation on each of the subgraphs separately, generating in total $Q$ augmentations for each of the $N$ subgraphs. For the probability matrix $M_i$ of each subgraph, we will set a threshold, and augment the subgraph by adding number of edges that have probabilities higher than this threshold. This threshold, as a hyperparameter of AugInf, will be discussed in the experimental section of this paper.

\subsection{Representation Learning}
AugInf will primarily use a GAE to learn a representation of the graph via a transformation of the graph structure into a low-dimensional latent space. 
Additionally, consistent with DeepInf,  AugInf will also use DeepWalk~\cite{perozzi2014deepwalk} to generate features of both the original graph $G$ and $Q$ augmented graphs as extra features for the social influence prediction.

\subsection{Neural Network Model} \label{section:joint}

We have implemented two variants of AugInf based on GCN and GAT, denoted by AugInf-GCN and AugInf-GAT. In our joint training model, AugInf-GCN uses GCN layers to obtain the output, while AugInf-GAT uses multi-head GAT layers.

\paragraph{Jointly trained model with train- and test-time augmentation}
We utilize joint training of both the GAE and the GNN prediction module to achieve two objectives: (1) obtain a more effective latent feature representation of the input graphs, (2) using these representations with the GNN prediction module to more accurately predict social influence. We experiment with the GNN prediction module implemented as either a GAT or a GCN for the social influence prediction.

The train-time augmentation process is shown in  Figure~\ref{method} (a). The inputs are the subgraphs sampled from graph $G$, along with their augmentations. We utilize the learned representation $Z$ as one of the GNN prediction module features, along with influence features and pretrained DeepWalk embeddings. The influence features are provided by the authors of DeepInf~\cite{qiu2018deepinf}, which record the user behaviour status (i.e. have they taken an action) and whether the user is the ego~\cite{qiu2018deepinf} user. The output of the GNN prediction module is a 2-dimension representation for each user (each node), corresponding to the negative log-likelihood. The test-time augmentation process is shown in Figure~\ref{method} (b).
For each test graph, we generate a number of augmented versions of the graph, learning representations and producing influence predictions for each, with the average of these predictions forming the final social influence prediction.

\paragraph{Loss}
The loss function of the joint model consists of two parts:
\begin{equation}
    L=L_D+L_{GAE},
\end{equation}
where $L_D$ is the loss of the final prediction and the ground truth, which is calculated by a negative log-likelihood, and $L_{GAE}$ is the loss of GAE which is calculated by the cross entropy of the decoded reconstructed graphs and the input graphs.

\section{Experiments}
\subsection{Datasets}
We evaluate using four datasets across different social network domains, namely OAG\footnote{OAG dataset details: www.openacademic.ai/} (Open Academic Graph), Digg\footnote{Digg dataset details: www.isi.edu/~lerman/downloads/digg2009.html/},  Twitter\footnote{Twitter dataset details: snap.stanford.edu/data/higgs-twitter.html} and Weibo\footnote{Weibo dataset details: www.aminer.cn/influencelocality}.
\begin{itemize}

\item The OAG graph consists of academic graphs representing the \textit{co-author} network in which the \textit{citation behaviours}, which we are predicting, are defined as the influence action behaviour.
\item The Digg dataset contains the timestamped voting behaviours of users on  \textit{stories} on a social news aggregation website. The edges of Digg graph are defined as \textit{following relationships} and the influence actions, which we are predicting, are \textit{voting behaviours}.
\item The Twitter dataset has been built by collecting Twitter data corresponding to tweets collected before, during and after the announcement of the discovery of the Higgs boson in 2012.  The graph is defined as a \textit{friendship} network, and the social action, which we are predicting, is defined as whether a user \textit{retweets} Higgs boson tweets.
\item The Weibo graph was built from 100 randomly selected users and their followers and followees. The social network is defined as a \textit{friendship} network, and the social action, which we are predicting, is defined as \textit{retweeting} behaviors in the Weibo social network.
\end{itemize}

These datasets were used previously by Qiu et al.~\cite{qiu2018deepinf}. Qiu et al.~\cite{qiu2018deepinf} sampled the entire social network into sub-networks with 50 nodes in each sub-network by using a random walk with restart, extracted features for each node and provided a ground-truth for the dataset. The statistics of the three datasets are shown in Table~\ref{Stat_data}.
\begin{table}[h]
\centering
\caption{The statistics of the datasets. $|V|$ and $|E|$ are the total numbers of nodes and edges of the original  dataset respectively and $N$ is the number of subgraphs after preprocessing. }
\resizebox{\linewidth}{!}{
\begin{tabular}{l|l|l|l|l}
\hline
  & OAG       & Digg      & Twitter   &Weibo \\ \hline
$|V|$ & 953,675   & 279,630    & 456,626  & 1,776,950  \\ \hline
$|E|$ & 4,151,463 & 1,548,126 & 12,508,413 & 308,489,739\\ \hline
$N$ & 499,848   & 24,428    & 362,888   & 779,164 \\ \hline
\end{tabular}}
\label{Stat_data}
\end{table}

\subsection{Evaluation Metrics}
We analyze several hyperparameters in our model and study how different hyperparameters may affect prediction performance. The performance is evaluated in terms of Area Under Curve (AUC) and $F_1$ score ($F_1$).
We compare our model with the state-of-the-art DeepInf~\cite{qiu2018deepinf} and PSCN~\cite{niepert2016learning} algorithms.

\subsection{Experiment Setup}

In our experiments we apply three augmentations to each graph with the augmentation hyperparameter threshold value set to 0.8 and train for 500 epochs. We will discuss the performance of varying these parameters in a later section. For the GAE component, each of the two hidden layers contain 64 hidden units for Digg and Twitter, and 32 for OAG and Weibo. They are trained with the Adagrad optimizer, using a 0.2 learning rate for OAG and Weibo, 0.05 for Digg and 0.1 for Twitter. Weight decay is set to 0.0005 all datasets except Digg, where it is 0.001. Additionally, we use dropout rate of 0.2. For the GNN prediction module, the first and second layers each contain 128 hidden units and the third layer, as the output layer, has two hidden units. There are eight attention heads in each GAT layer, which means each head needs to process 16 hidden units for Digg and Twitter,  with four attention heads for OAG and Weibo, which means each head needs to process 32 hidden units. The nonlinear activation function we use for both augmentation and prediction ($\sigma$ in Eq.~\ref{gcn} and~\ref{hi}) is the exponential linear unit (ELU)~\cite{clevert2015fast}.

\subsection{Experimental Results}
We report the performance of our method (AugInf-GAT and AugInf-GCN) over ten runs. The mean and standard deviations are shown in Table~\ref{DeepInf_result}. For clarity, the input features of our models are influence features and pretrained DeepWalk embeddings, consistent with the end-to-end method DeepInf~\cite{qiu2018deepinf}, but without any hand-crafted vertex features. As an experimental comparison, we used the state-of-the-art GNN methods DeepInf and PSCN~\cite{niepert2016learning} as the baselines.

\begin{table}[h]
\tiny
\centering
\caption{The performance of two AugInf models on different datasets, along with the performance of the baselines without vertex features.}
\resizebox{\linewidth}{!}{
\begin{tabular}{l|l|l|l}
\hline
Dataset                        & Model & AUC    & $F_1$    \\\hline
\multirow{5}{*}{Digg} & DeepInf-GAT   & 0.8882($\pm$0.011) & 0.7052($\pm$0.010) \\\cline{2-4}
                 
                      & DeepInf-GCN   & 0.8372($\pm$0.007)  & 0.6404($\pm$0.006) \\\cline{2-4}
                      & PSCN          & 0.8499($\pm$0.013) & 0.6636($\pm$0.011) \\\cline{2-4}
                      & AugInf-GCN    & 0.8580($\pm$0.013)  & 0.6711($\pm$0.024) \\\cline{2-4}
                      & AugInf-GAT    & \textbf{0.9067}($\pm$0.011) & \textbf{0.7385}($\pm$0.017)\\\hline
                      
\multirow{5}{*}{Twitter} & DeepInf-GAT  & 0.7843($\pm$0.002) & 0.5484($\pm$0.002) \\\cline{2-4}
                         & DeepInf-GCN  & 0.7615($\pm$0.004) & 0.5271($\pm$0.004) \\\cline{2-4}
                         & PSCN         & 0.7664($\pm$0.004) & 0.5319($\pm$0.006) \\\cline{2-4}
                         & AugInf-GCN   & 0.7694($\pm$0.003) & 0.5360($\pm$0.006) \\\cline{2-4}
                         & AugInf-GAT   & \textbf{0.7861}($\pm$0.004) & \textbf{0.5486}($\pm$0.007) \\\hline
                         
\multirow{5}{*}{OAG} & DeepInf-GAT  & 0.6814($\pm$0.006) & 0.4544($\pm$0.004) \\\cline{2-4}
                     & DeepInf-GCN  & 0.6281($\pm$0.002) & 0.4231($\pm$0.003)  \\\cline{2-4}
                     & PSCN         & 0.6556($\pm$0.006) & 0.4372($\pm$0.029) \\\cline{2-4}
                     & AugInf-GCN   & 0.6334($\pm$0.003) & 0.4229($\pm$0.011) \\\cline{2-4}
                     & AugInf-GAT   & \textbf{0.6889}($\pm$0.012) & \textbf{0.4602}($\pm$0.004) \\\hline
\multirow{5}{*}{Weibo} & DeepInf-GAT  & \textbf{0.8212}($\pm$0.003) & 0.5770($\pm$0.003) \\\cline{2-4}
                      & DeepInf-GCN  & 0.7706($\pm$0.002) & 0.5312($\pm$0.004) \\\cline{2-4}
                      & PSCN         & 0.8012($\pm$0.004) & 0.5625($\pm$0.002)\\\cline{2-4}
                      & AugInf-GCN   & 0.7531($\pm$0.006) & 0.5147($\pm$0.004)\\\cline{2-4}
                      & AugInf-GAT   & 0.8124($\pm$0.005) & \textbf{0.5785}($\pm$0.007) \\\hline
\end{tabular}}
\label{DeepInf_result}
\end{table}

Our proposed method AugInf-GAT achieves better performance over most of the baselines in terms of AUC. However, consistent with all models, the GCN-based approaches do not perform well, as GCN uses the unweighted aggregations over the neighbours’ representations when calculating a node's representations, a mechanism that appears not to be suitable for tasks like social influence prediction, which benefits from considering the importance of neighbouring nodes.

Referring to the statistics of the datasets in Table~\ref{Stat_data}, we can see that the performance improvement of AugInf-GAT is particularly clear on the datasets with much fewer edges (Digg) but limited on datasets with more edges (Twitter and Weibo). We believe that this is because the Twitter and Weibo datasets contain enough edges to learn a sufficiently comprehensive representation, hence less benefit is gained from the augmentation. We will further investigate the effect of removing edges from graphs as part of data augmentation in future work. Nonetheless, particularly for smaller graphs, we believe our proposed approach of train- and test-time augmentation can provide additional performance.

\subsubsection{Hyperparameter Analysis}
We conduct an hyperparameter analysis on the Digg dataset with the same hyperparameters values mentioned previously, unless stated otherwise.

\paragraph{Number of Augmentations}
For the data augmentation, we analyze the effect of the number of augmentations on the performance of AugInf. We successively apply one to eight augmentations while leaving the other parameters constant. The results of this are shown in Figure~\ref{num}. When we apply four augmentations, the highest performance is achieved. After that, as the number of augmentations increase, the AUC score stabilises. Interestingly, the $F_1$ score does not appear to be affected by the number of augmented graphs and remains stable between 0.73 and 0.74.

\begin{figure}[h]
  \centering
  \includegraphics[width=\linewidth]{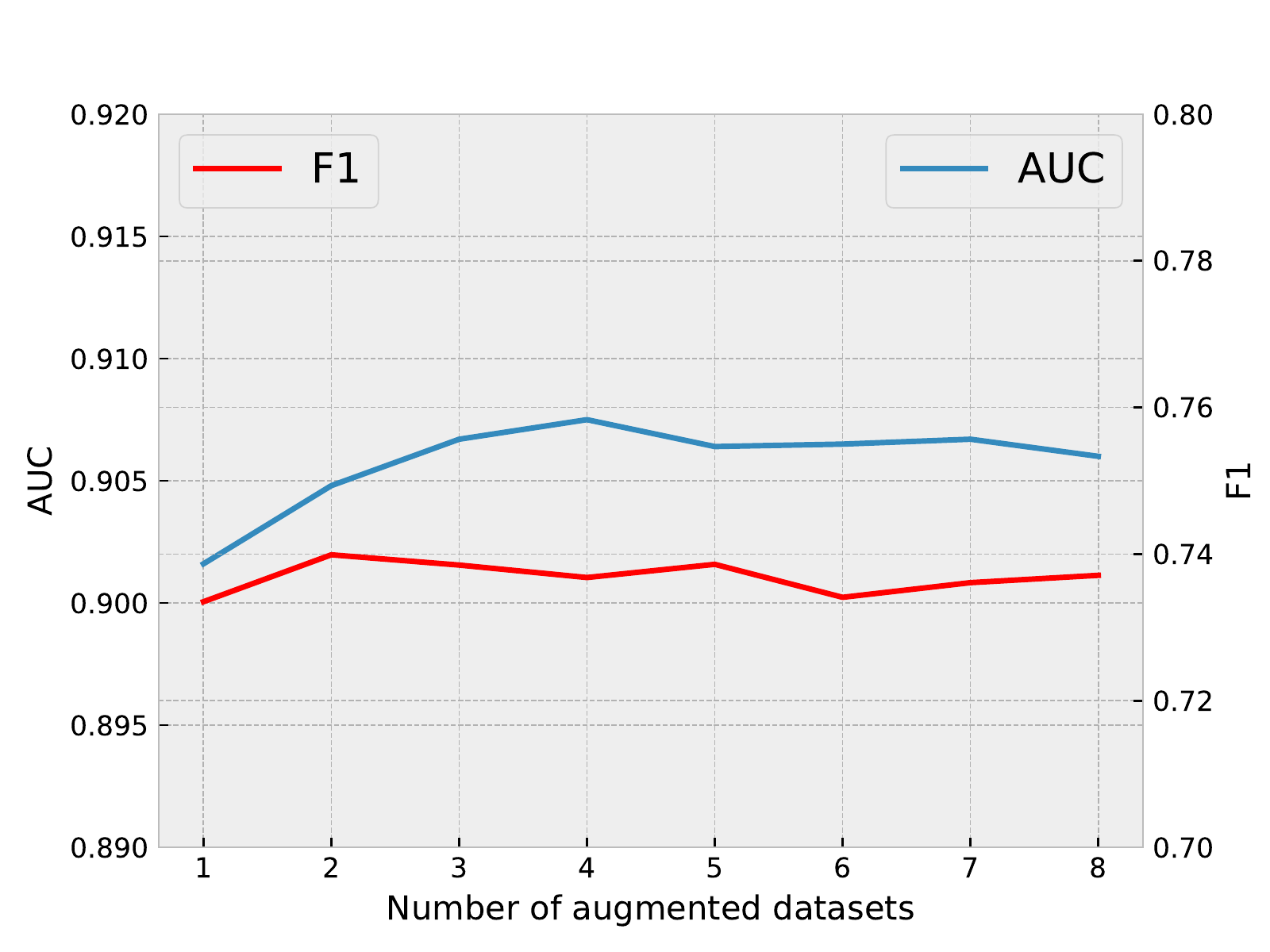} 
  \caption{The performance of AugInf-GAT on Digg as we vary the number of augmentations. The value of threshold is set to 0.8.}
  \label{num} 

\end{figure}

\paragraph{The Threshold for Augmentation} Another parameter we analyze is the threshold that determines which edges may be added. The results of this are shown in Figure~\ref{edge}. When the threshold is set to 0.8 for Digg dataset, our method achieves the highest performance, while on average the number of edges per dataset increases by $2.7\%$. As we increase the number of added edges, the performance of our method decreases.

\begin{figure}[h]
  \centering
  \includegraphics[width=\linewidth]{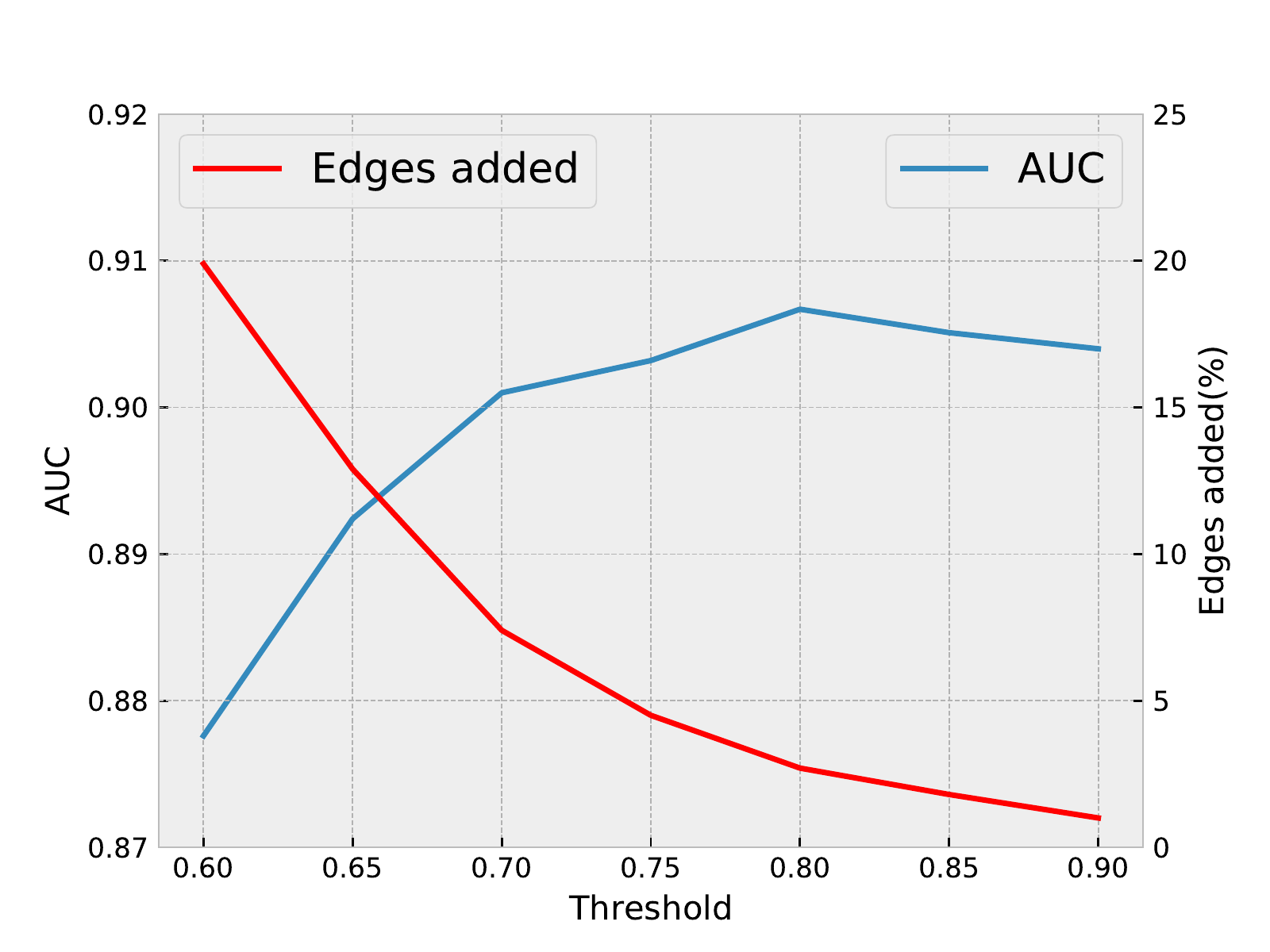} 
  \caption{The performance of AugInf-GAT on Digg as we vary the parameter that controls the minimum quality of edges that are added for augmentation. The number of augmentations is set to three.}
  \label{edge} 
\end{figure}

\subsection{Ablation Study}

We have further evaluated AugInf-GAT and AugInf-GCN on the individual components of our approach, to determine the contribution of each component to the overall performance. 

There are three main components in our approach: (1) train-time augmentation (2) test-time augmentation and (3) the jointly trained model. We have evaluated the following combinations on the Digg dataset:
\begin{itemize}
    \item Ablation $\#1$: The method contains only a GNN prediction module. We use this set of experiments as a baseline to compare with other combinations when other components are added.
    \item Ablation $\#2$: The method contains the jointly trained model with no augmentation. The purpose of this experiment is to measure the effect of joint training without augmentation.
    \item Ablation $\#3$: The method uses a GNN prediction module with train-time augmentation only. The purpose of this experiment is to measure the effect of train-time augmentation.
    \item Ablation $\#4$: The method uses a GNN prediction module with test-time augmentation only. The purpose of this experiment is to measure the effect of test-time augmentation.
    \item Ablation $\#5$: The method uses a GNN prediction module with both train-time and test-time augmentation. The purpose of this experiment is to measure the effect of both forms of augmentation together.
    \item Ablation $\#6$: The method contains the joint training model with train-time augmentation only. The purpose of this experiment is to evaluate the extent of which using train-time augmentation improves the joint training model.
    \item Ablation $\#7$: The method contains the joint train-model with test-time augmentation only. The purpose of this experiment is to evaluate the extent of which using test-time augmentation improves the joint training model.
    \item Complete $\#8$: The method is our complete method, AugInf, consisting of all components.
\end{itemize}

The results with the GAT prediction module are shown in Figure~\ref{each_gat}. Through the comparison with $\#1$ and $\#2$, we can see the jointly trained model contributes to the performance improvement. Through the comparison among the ablations $\#1$, $\#3$, $\#4$ and $\#5$, we observe that the augmentations can improve the GAT performance, but there is little difference between using train-time augmentation and test-time augmentation separately or together, without the jointly trained model. The result of version $\#8$, the complete AugInf method, shows using train- and test-time augmentation, along with the jointly trained model, has the highest AUC score and $F_1$ score.

\begin{figure}[h]
  \centering
  \includegraphics[width=\linewidth]{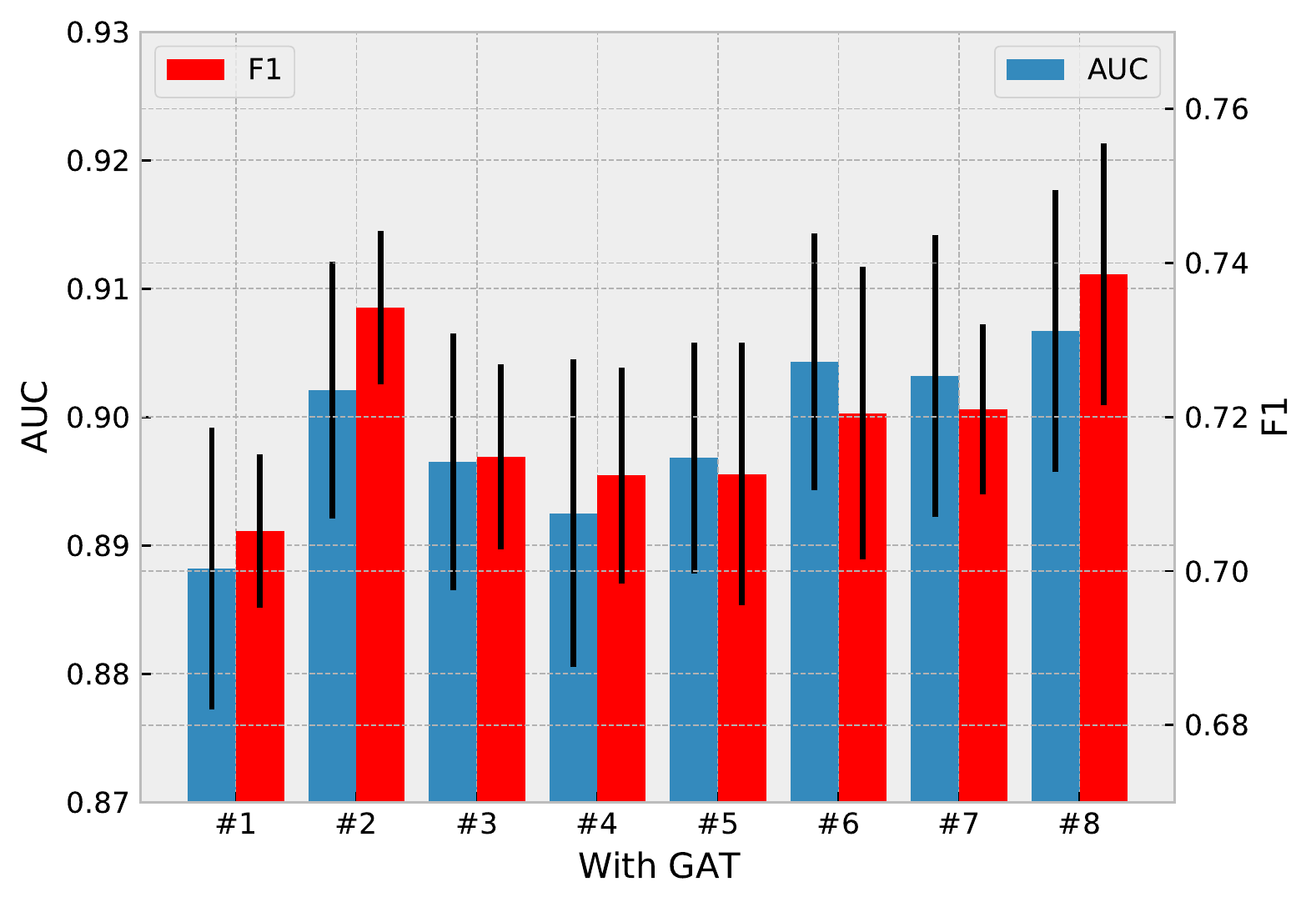} 
  \caption{The performance of the ablation study using GAT where $\#8$ is our complete method consisting of all components, AugInf.}
\label{each_gat}
\end{figure}

The results with the GCN prediction module are shown in Figure~\ref{each_gcn}. The AUC scores of this set of experiments are between 0.83 and 0.86. 
We can see that all components have contributed to some extent to the performance of the GCN prediction module, but overall, AugInf when using a GCN for social influence prediction achieves lower performance than AugInf when using a GAT for social influence prediction.

\begin{figure}[h]
  \centering
  \includegraphics[width=\linewidth]{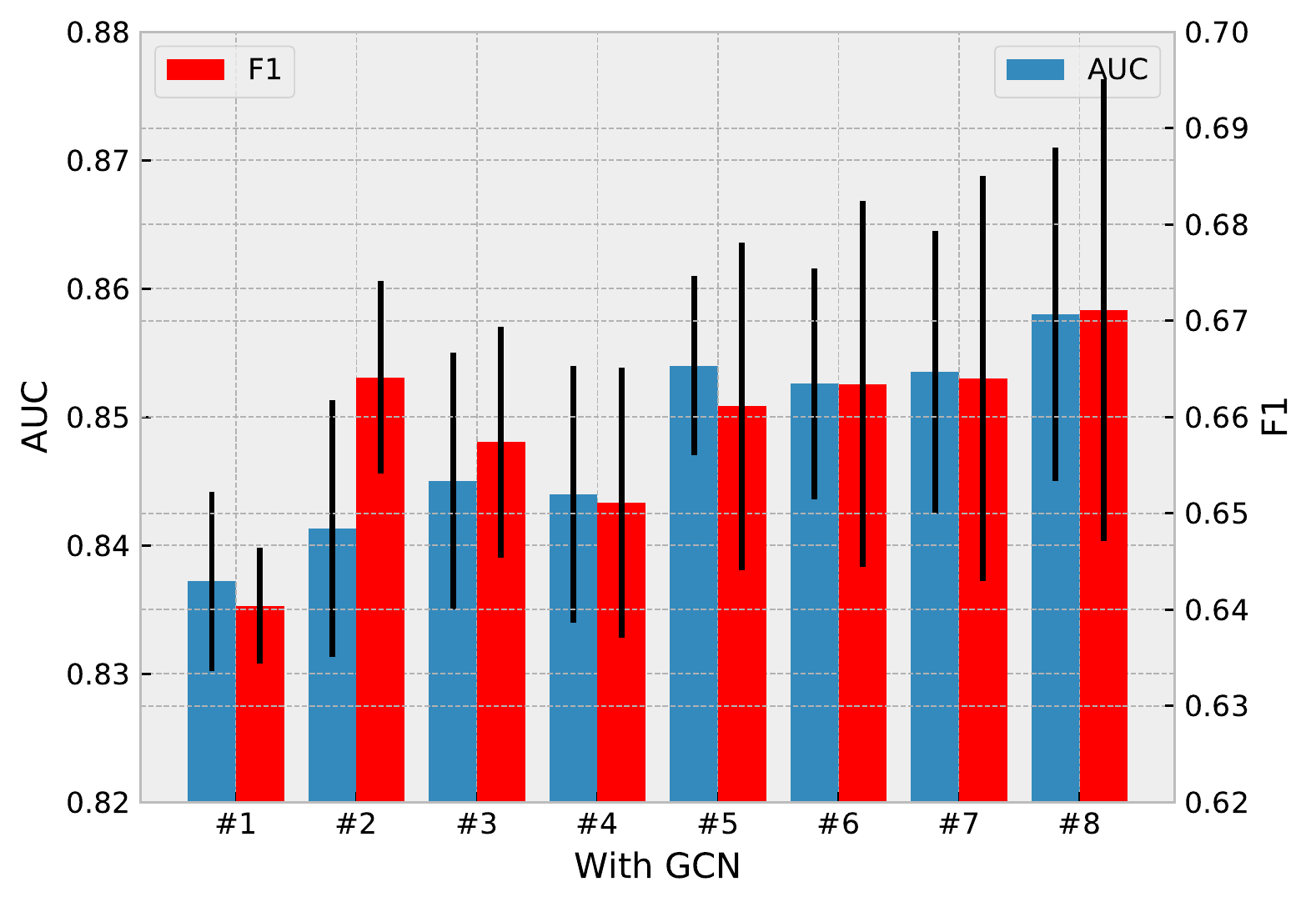} 
  \caption{The performance of using different combinations of components with GCN, where $\#8$ is our complete model consisting of all components, AugInf.}
\label{each_gcn}
\end{figure}

\section{Conclusions}
In this paper we  proposed a new social influence prediction method, AugInf, which incorporates train- and test-time augmentation with a jointly trained graph neural network approach. During training, this method takes into account the losses of both the graph representation learning and downstream social influence prediction task. We improve performance by applying numerous augmentations to the graphs using variational graph auto-encoders at both train- and test-time. Via an ablation study we show that the jointly trained model obtains more effective latent feature representations by using the joint loss along with both the train- and test-time augmentations.
We compare our proposed end-to-end method with the state-of-the-art on several social network datasets. The experimental results show that our proposed method, AugInf-GAT, can improve the performance of predicting social influence on a number of social networks, and in particular, on the smallest of the social network graphs.

\bibliographystyle{IEEEtranN}
\bibliography{references}

\end{document}